\documentclass[letterpaper, 10 pt, conference]{ieeeconf} 
\IEEEoverridecommandlockouts             
\usepackage{amsthm}
\usepackage{amsmath}
\usepackage{amssymb}
\usepackage{mathtools, nccmath}

\usepackage{cite}
\usepackage{microtype}
\usepackage{graphicx}
\usepackage{subfigure}
\usepackage{booktabs} 
\usepackage{array}
\usepackage{algorithm}
\usepackage{algpseudocode}
\algrenewcommand{\algorithmiccomment}[1]{\hskip.3cm $/*$\texttt{#1}$*/$}
\usepackage{adjustbox}
\usepackage{hyperref}
\usepackage{import}
\usepackage{sjmacros}
\usepackage{ulem}
\usepackage{multirow}

\makeatletter
\newcommand\subsubsubsection{\@startsection{paragraph}{4}{\z@}%
    {-3.25ex\@plus -1ex \@minus -.2ex}%
    {1.5ex \@plus .2ex}%
    {\normalfont\normalsize\bfseries}}
\makeatother
\makeatletter
\newcommand{\tpmod}[1]{{\@displayfalse\pmod{#1}}}
\makeatother

\usepackage[capitalize,noabbrev]{cleveref}

\theoremstyle{plain}
\newtheorem{theorem}{Theorem}[section]

\newtheorem{lemma}[theorem]{Lemma}

\theoremstyle{definition}
\newtheorem{definition}[theorem]{Definition}

\theoremstyle{remark}
\newtheorem{remark}[theorem]{Remark}

\title{\LARGE \bf
Efficient Replay Memory Architectures in Multi-Agent\\ Reinforcement Learning for Traffic Congestion Control}

\author{Mukul Chodhary$^{1}$, Kevin Octavian$^{2}$, SooJean Han$^{2}$%
\thanks{$^1$Department of Electrical and Electronic Engineering, University of Melbourne, Australia. $^2$School of Electrical Engineering, KAIST, S. Korea.}
\thanks{Corresponding author: \texttt{soojean@kaist.ac.kr}.}
}

\begin{document}
\maketitle
\thispagestyle{plain}
\pagestyle{plain}

\begin{abstract}
    Episodic control, inspired by the role of episodic memory in the human brain, has been shown to improve the sample inefficiency of model-free reinforcement learning by reusing high-return past experiences. 
However, the memory growth of episodic control is undesirable in large-scale multi-agent problems such as vehicle traffic management.
This paper proposes a novel replay memory architecture called Dual-Memory Integrated Learning, to augment to multi-agent reinforcement learning methods for congestion control via adaptive light signal scheduling.
Our dual-memory architecture mimics two core capabilities of human decision-making.
First, it relies on diverse types of memory--semantic and episodic, short-term and long-term--in order to remember high-return states that occur often in the network and filter out states that don't.
Second, it employs equivalence classes to group together similar state-action pairs and that can be controlled using the same action (i.e., light signal sequence).
Theoretical analyses establish memory growth bounds, and simulation experiments on several intersection networks showcase improved congestion performance (e.g., vehicle throughput) from our method.
\end{abstract}



\section{Introduction}
Like many large-scale network applications, traffic management is often characterized by a high degree of spatiotemporal symmetries.
For example, some metropolitan intersection networks exhibit spatial symmetries from a grid-like topology, and human routine gives rise to temporal symmetries such as morning and evening commute times.
Multiagent reinforcement learning (MARL) algorithms have demonstrated potential to develop effective policies for traffic management problems.
However, general reinforcement learning is still often hindered by \textit{sample inefficiency}, requiring a significant amount of training data to learn a task that humans could learn from fewer exposures~\cite{mnih_human-level_2015, haarnoja_soft_2018}.
A key attribute in the efficiency of human decision-making comes from the ability to use diverse types of memory to learn and recall patterns in the environment.
\textit{Memory}, then, becomes an important aspect of MARL (and RL in general) to effectively store past experiences while exploring the environment, especially in settings which exhibit many symmetries.

A common way of implementing memory is through experience replay buffers~\cite{fedus20a}, which store past experiences (i.e., tuples of state, action, reward, and next state) in a tabular fashion, then supplements the agent's training process on randomly-sampled mininbatches from the buffer.
Another method is neural episodic control~\cite{pritzel17}, which implements a differentiable neural dictionary (DND), whose differentiability allows for end-to-end learning to be achieved, in contrast to the replay buffer.
%
However, most memory tables (e.g., replay buffers, DNDs) are structured such that every new experience encountered is stored as a new entry, potentially leading to linear memory growth.
A rapidly growing memory table is undesirable in applications such as large-scale networks, where the state-space is large, and also unnecessary in traffic management, where natural spatial and temporal repetition can be leveraged. 

This paper proposes \textit{dual-memory integrated learning}, a novel memory architecture to include in MARL algorithms for efficient control and decision-making over large-scale networks.
We are largely motivated by the human brain, which effectively uses \textit{different combinations} of memory to make optimal decisions instead of just one type (like a replay buffer).
Prior work on MARL for vehicle traffic congestion control~\cite{pmlr-v211-han23a} has demonstrated that learning patterns can prevent the agent from spending unnecessary time and energy on redundantly developing adaptive light signal policies for scenarios that have been previously observed.
The pattern learning was largely implemented based on episodic memory~\cite{gershman_reinforcement_2017}; in contrast to semantic memory, which stores only the statistical summaries of event trajectories, episodic memory saves episodic traces along with their returns.
Moreover, recent work in incorporating symmetry inductive biases into conventional neural networks has demonstrated the powerful idea of encoding symmetries directly into the model (e.g., rotation~\cite{cohen16}, permutations~\cite{battaglia18}) to reduce the number of parameters that need to be trained.

Motivated by the above, our present paper differs from~\cite{pmlr-v211-han23a} by improving the table's rate of growth in two primary ways.
First, we develop a two-step short-term and long-term memory approach for a more robust way of identifying which patterns occur frequently enough in the system to warrant permanent storage in memory.
Second, we borrow ideas from group-equivariance to define the symmetry relations to form equivalence classes of patterns such that each equivalence class is assigned to a single action sequence.
%

Our paper is organized as follows.
We begin with a brief review of relevant background in~\sec{background}, such as MARL, memory in RL, and equivariance.
The main problem formulation is described in~\sec{problem_formulation}.
Our main contributions--the dual-memory integrated learning framework, equivalence classes and other constituents--are detailed in~\sec{memory}.
The remaining sections compare the performance of our dual-memory integrated learning framework against state-of-the-art baselines, with theoretical analyses in~\sec{analysis} and simulation experiments in~\sec{experiments}.
We conclude the paper in~\sec{conclusion}.



\section{Background}
\label{sec:background}
\subsection{Reinforcement Learning for Network Control}\label{subsec:network_rl}
General reinforcement learning (RL) problems involve an agent interacting with an environment which is typically represented as a Markov decision process (MDP) $(\mathcal{S}, \mathcal{A}, \mathcal{R}, \mathcal{P})$, where $\Scal$ is the state space, $\Acal$ is the set of actions, $\Rcal{\,\equiv\,}R_{\avect_t}(\xvect_{t}, \xvect_{t+1})$ is the reward function, and $\Pcal{\,\equiv\,}P_{\avect_t}(\mathbf{\cdot|x_t})$ is a probabilistic transition function.
The goal is to determine an optimal policy $\pi^*$ to maximize cumulative expected reward.
Methods for computing $\pi^*$ (e.g., Q-learning, SARSA) often involve iteratively optimizing the $Q$-function $Q^{\pi}(\svect,\avect) \triangleq \Ebb_{\pi}[\sum_t \gamma^t r_t | \svect,\avect]$, where $\svect$ is the initial state, $\avect$ is the initial action, and $\gamma\in(0,1)$ is a discount factor.

A class of RL methods designed specifically for scalable implementation is \textit{multi-agent RL (MARL)} methods, where the focus is on how to configure multiple agents to work together and achieve a shared goal.
In addition to the 4-elements tuple in single-agent MDP, MARL requires a set of observations $\Ocal$, which is the partial information obtained by each agent. 
Here, the observations simply mean that each agent in a multiagent setting only has a partial view of the entire joint state (e.g., a single intersection in the entire traffic network). 
Since there is no uncertainty affecting these partial observations, this is still considered a MDP (rather than a POMDP) despite the adjustment in notation; the main difference is in whether it is implemented in a multiagent way or a single-agent way.

\subsection{Reinforcement Learning with Memory}\label{subsec:memory_rl}
Including memory buffers in deep RL algorithms is a common way to improve sample inefficiency.
For example, \textit{deep $Q$-network (DQN)}~\cite{mnih2013atari} agents use experience replay to store state transitions $(\svect_t,\avect_t,r_t,\svect_{t+1})$, and train on minibatches of the replay buffer.
Although the simplest method samples the minibatches randomly, several other techniques have been proposed for more stable training, including prioritized replay \cite{schaul2016prioritized} and hindsight experience replay \cite{andrychowicz2018hindsight}.

One key attribute of human intelligence is our ability to effectively use different types of memory for efficient decision-making.
For example, \cite{tulving:episem} classified two subcategories of explicit memory: semantic memory and episodic memory. In contrast to \textit{semantic memory}, which encodes general world knowledge (e.g., the rules of the road, how to drive a car), \textit{episodic memory} also adds contextual information which makes each experience highly personal (e.g., going on a road trip).
The heterogeneous human memory system suggests that machine intelligence can also benefit from a diverse implementation of memory architectures. 
However, conventional machine learning models are trained on data analogous to experiences recorded by semantic memory.
The class of \textit{episodic control} methods ~\cite {lengyel07,blundell16,pritzel17} leverage episodic memory to enhance training efficiency. 
This approach involves recalling specific instances of highly rewarding experiences, thereby accelerating the learning process. 
By prioritizing memorization and storage of impactful experiences, episodic control methods generally outperform traditional DQNs equipped with experience replay in terms of learning efficiency.

Previous work~\cite{pmlr-v211-han23a} introduced a controller architecture based on \textit{pattern-learning with memory and prediction (PLMP)}, which extends episodic control by identifying recurrent ``patterns'' in the environment.
Compared to other control methods, PLMP reduces computation time and redundancy by memorizing past patterns and predicting future patterns.
Furthermore, a good choice of pattern embedding creates a control pipeline that can be trained end-to-end; see, e.g., the \textit{differentiable neural dictionary (DND)} in neural episodic control (NEC)~\cite{pritzel_neural_2017}.

\subsection{Group Equivariance and Invariance}\label{subsec:equivariance}
Equivariant/invariant neural network architectures and symmetry inductive biases were developed with the premise that humans typically perform very well on classification tasks regardless of the orientation and position of the object they view.
These methods are largely based on concepts from group theory, where a group $(\Gcal,\circ)$ is a set of transformations $\Gcal$ that are applied via operation $\circ$ to objects like the dataset $\Xcal$, the set of labels $\Ycal$, or each layer of the neural network architecture.
A function $F{\,:\,}\Xcal{\,\to\,}\Ycal$ is \textit{invariant} if $F(g\circ\xvect){\,=\,}F(\xvect)$, i.e., both input $\xvect$ and transformed input $g\circ\xvect$ return the same label $y$.
$F$ is \textit{equivariant} if $F(g\circ\xvect) = g\circ F(\xvect)$, i.e., if input $\xvect$ returns label $y$, then transformed input $g\circ\xvect$ returns transformed label $g\circ y$.

Prior work such as PLMP and NEC suggests the pattern embedding method can also be optimized.
We leverage invariance and equivariance to abstract the state-space into \textit{equivalence classes} of patterns.
The main advantage of equivalence classes is efficient representation, which limits the rate of growth in the number of entries in the memory table.
All patterns in the same equivalence class are then controlled using slight variations to the original policy.

\section{Problem Formulation}
\label{sec:problem_formulation}
Our methods in this paper are developed specifically to mitigate congestion problems (e.g., maximizing throughput, minimizing journey time) in large-scale traffic networks.
We consider settings where the vehicles moving throughout the network are discrete counts over discrete time, instead of continuous traffic flows.
Because routing is not the focus of our paper, each arrival has a predetermined final destination and simply follows the shortest path to reach it.

The generic vehicle intersection network is represented as a graph $\Gcal{\,\triangleq\,}(\Ncal, \Ecal)$, with the set of nodes $\Ncal$, and directed edges $\Ecal$ that connect between two nodes $n,m{\,\in\,}\Ncal$.
We distinguish between two types of nodes $\Ncal{\,\triangleq\,}\Ncal_F\cup\Ncal_I$: (i) fringe nodes $\Ncal_F$ and (ii) intersection nodes $\Ncal_I$.
Fringe nodes lie on the edges of the network and are used as a medium for two tasks: 1) feeding incoming arrivals into the network, and 2) enabling outgoing departures to exit the network entirely after completing their route.
Incoming traffic can be modeled either from real-world data or synthetically-generated arrival processes of common distributions, e.g., compound Poisson processes.
Each intersection implements the dynamics for local conflict resolution among multiple directions of traffic.
A controller (MARL agent) is implemented at each intersection to mitigate congestion locally, and then the overall network's congestion levels are optimized by coordinating these local controllers via a \textit{central unit (CPU)}.

We now define the MARL MDP for congestion control of our traffic network.\\
%
\textbf{State}: A node state is represented with the vector $\mathbf{x}_t{\,\equiv\,}[x_t(j)]$, where $j$ denotes the orientation of incoming traffic and each element $x_t(j){\,\in\,}\mathbb{Z}^{\geq 0}$ is the amount of traffic present at time $t$.
For example, in a 4-way intersection with 3 lanes, the indices are defined as $(j){\,\triangleq\,}(D/i)$, where $D{\,\in\,}\{E,N,W,S\}$ is the direction, and $i{\,\in\,}\{\texttt{rt},\texttt{fwd},\texttt{lft}\}$ is the lane.\\
\textbf{Action}: $\mathcal{A}$ is the set of actions taken by each traffic light.
One example design involves splitting all the incoming traffic into straight-line motion, adjacent motion, and a combination of both.
Assuming right-turns are permitted whenever, this yields $\abs{\Acal}{\,=\,}8$ possible modes: $\mathcal{A} = \{(E/W, F), (E/W,L), (N/S, F), (N/S, L), (E, F/L),\\ (N, F/L), (W, F/L), (S, F/L)\}$.\\
\textbf{Transition Function}: At each intersection, $\Pcal$ can be derived from flow conservation dynamics with two main flows: all incoming items increase the counts recorded in the state $\xvect_t$, while all departing items decrease them. The rate of departing items is dependent on the action applied at the intersection.
For concrete modeling, we specify two additional quantities: (i) a baseline duration $t_a{\,\in\,}\Nbb$ over which a single action is applied, and (ii) the maximum rate $v^{*}{\,\in\,}\Nbb$ of items per orientation which can depart the node in one timestep when the current action allows them to pass, i.e., $x_{t+1}(j){\,-\,}x_t(j){\,=\,}\min(x_t(j),v^*)$ if action $\avect_t{\,\in\,}\Acal$ allows traffic in orientation $j$ to pass.\\ 
%
\textbf{Reward Function}: An agent aims to reduce the congestion at each orientation of its intersection. 
\begin{align}\label{eq:default_reward}
    R_{\avect}(\xvect_{t}, \xvect_{t+1}) &= \sum_{j} r(x_{t}(j), x_{t+1}(j)), \text{ where}\\ 
    r(x, y)&\triangleq \begin{cases}
        1, & \text{if } y - x \leq \alpha \\
        -1, & \text{if } y - x \geq \alpha + \theta \\
        0 & \text{else}
        \end{cases}\notag
\end{align}
Hyperparameters are added to limit the rate of congestion per timestep.
$\alpha$ is a relaxation parameter with $\alpha{\,=\,}0$ meaning reward $r$ is only given when the flow is conserved; we want vehicles to pass through intersection $j$ without interruption from the light signal.
$\theta$ is the maximum allowed addition of vehicles at the next timestep for which penalty is not given.
To avoid the case where the agent sticks to one action only, we penalize the agent when the number of cars increases over some $\alpha + \theta$.

\begin{remark}[Q-Value Function Design]\label{rmk:marl_avg}
In general congestion control problems, where flow conservation must hold, the $Q$-function of the entire problem cannot be decomposed into a direct sum of the agents' individual $Q$-functions. 
We thus modify the MARL decomposition from Section~\ref{subsec:network_rl} into an averaging of rewards instead: $Q(\Scal, \Acal) \approx (1/N) \sum_{i \in \mathcal{I}} \hat{Q}_i(s^i, a^i)$. This is to ensure the scale of Q-function in centralised memory is similar to the individual Q-functions.
\end{remark}


\section{Dual-Memory Integrated Learning}\label{sec:memory}

\subsection{Equivalence Classes}\label{subsec:equivalence}
To mitigate the rate of growth in our $Q$-tables, we use invariance and equivariance to implement the equivalence classes discussed in Section~\ref{subsec:equivariance}.
We implement two different types of equivalence class embeddings.
First, we define a simple embedding method that considers a shift-invariance of 1 radial distance, similar to~\cite{pmlr-v211-han23a}.
Second, we define a complex embedding method that adds two more definitions of equivalence to the simple embedding: rotational equivariance and scale invariance.
\begin{figure}[H]
    \centering
    \includegraphics[width=0.45\textwidth]{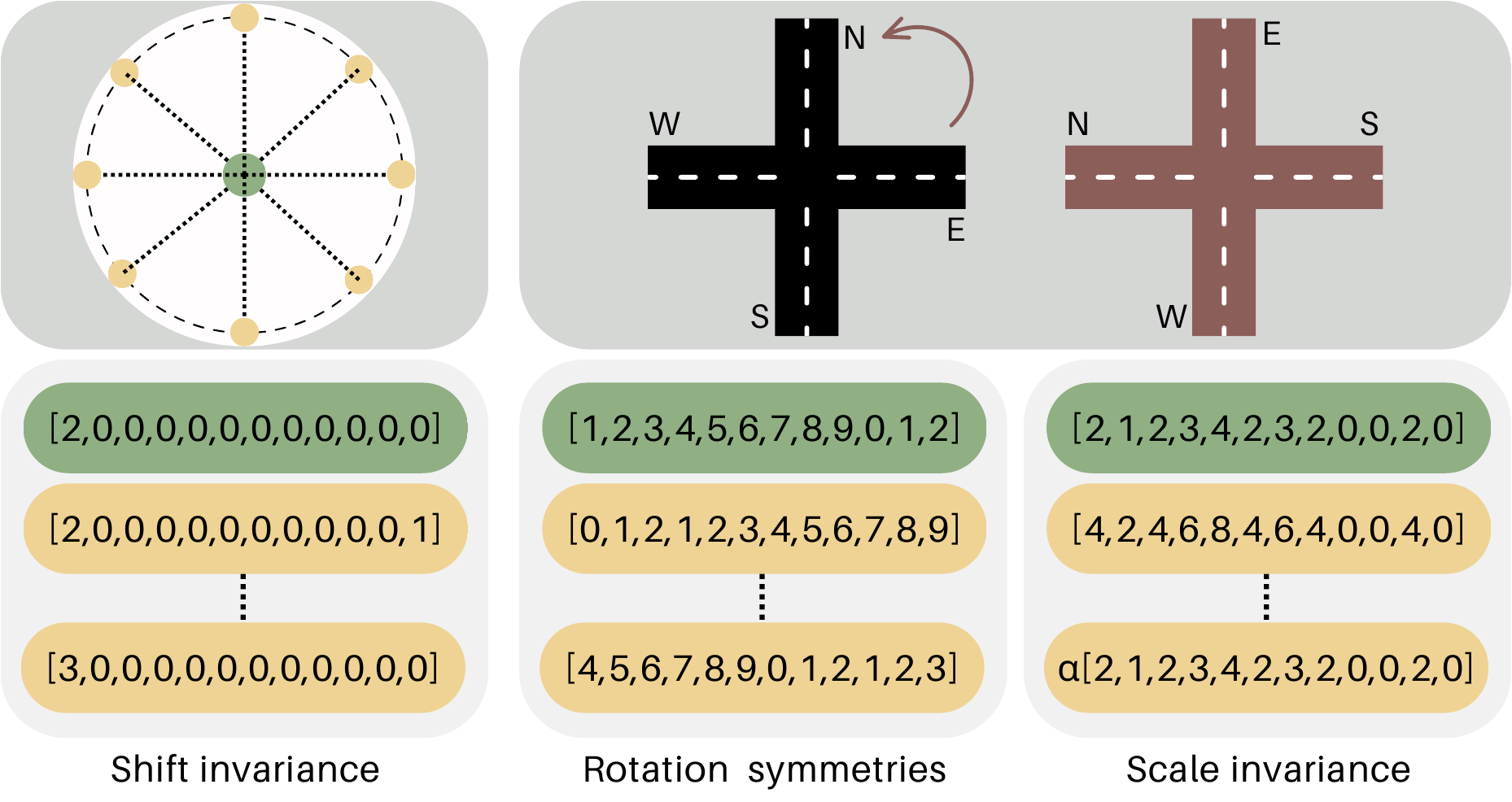}
    \label{fig:eq_classes}
    \caption{Two equivalence class types implemented for traffic control.
    }
    \label{fig:gs_eq_c}
\end{figure}
\vspace{-.4cm}
Figure~\ref{fig:gs_eq_c} visualizes at least two equivalent states (yellow) for a sample ``key'' state (green) in our vehicle traffic system.
Each new state which is observed at the intersection is compared with the existing key states to see if it can be added to an existing equivalence class.
Otherwise, a new key state is created.
A single traffic light is assigned to control all the states within a single equivalence class. 
Looking up $Q$-values amounts to looking at only the ``key'' state representing the equivalence class rather than all entries inside it.
For the equivariant equivalence embeddings, a transformed state is controlled using the \textit{transformed} light signal, while for invariant embeddings, the same light signal is used.

\subsection{The Memory Architecture}\label{subsec:memory_architecture}
Our proposed memory architecture consists of two phases: \textit{short-term memory (STM)}, which learns directly from the environment, and \textit{long-term memory (LTM)}, which consolidates past experiences. 
The dual memory architecture was designed with modularity in mind, aiming for enhanced control over memory functionality and their mutual interaction. The separation into short-term and long-term memories serves the purpose of further minimizing memory usage. 
The STM selectively retains frequently occurring states in the environment, discarding those that occur infrequently, as they hold little value for long-term retention.
%
Each intersection stores their own high quality states-action pair in the long-term memory. To allow information exchange between intersection, we also implemented a CPU, which aggregates the high quality states-action pair of all intersections and feed them back to each intersection. This information-exchanging process between the agents' memories and the CPU is dictated by a \textit{trigger rule}, which is described further later. 

(\textbf{Short Term Memory (STM)}) [\alg{stm}] Initially, all states are inserted in the STM, making it behave like a typical Q-table~\cite{suttonbarto} (see also,``SARSA Agent'' in Section~\ref{subsec:agent_types}). After some duration of time, the STM is allowed to only learn local temporal information rather than long-term information. The most significant data is then stored until a staging rule moves it to the LTM. 
Below, $S', A', S, A$ is the abstracted state, abstracted action, original state, and original action respectively.

\begin{algorithm}
\small{
\caption{Inserting into Short-Term Memory}
\label{alg:stm}
\begin{algorithmic}[1]
\State{\textbf{Short-Term Memory:} $\Mcal$}
\State{\textbf{Input:} State $S$, Action $A$, Q-value $Q$}
   
    \\ \Comment{State Abstraction Rule}
\State{$S_{\text{set}} \leftarrow \text{getAbstractedStateSet}(S)$}
\State{$S' \leftarrow S_{\text{set}} \cap \mathcal{M}$}
\If {$S' \neq \varnothing$}
\State{ $A' \leftarrow \text{getEquivalentAction}(A, S, S')$ }
\Else
\Comment{New State Discovered}
        \State{$S', A' \leftarrow S, A$}
        \State{$\Mcal$.append($S'$)} 
\EndIf
    \\ \Comment{Update Rule}
    \State{$\Mcal[S', A'] \leftarrow Q$}
    \\ \Comment{LTM Staging Rule}
    \If{$\Mcal$ is full}
    \Comment{stage the $m_s/2$ states to LTM with the highest Q-values}
    \State{$\Mcal$.stageToLTM()}
    \State{$\Mcal$.empty()}
    \EndIf
\end{algorithmic}
}
\end{algorithm}

(\textbf{Long Term Memory (LTM})) [\alg{ltm}] The LTM table stores the best action so far for a state provided by the STM.
Because of the initial screening phase done by the STM, the overall size of LTM never exceeds that of a typical Q-table.
We formalize this via the analyses in Section~\ref{sec:analysis}.

\begin{algorithm}
\small{
\caption{Inserting into Long-Term Memory}
\label{alg:ltm}
\begin{algorithmic}[1]
\State{\textbf{Long-Term Memory:} $\Mcal$}
\State{\textbf{Input: } stagedStates}
\For{State, best action, best reward in stagedStates}
\State{$S' \leftarrow \text{getAbstractedStateFromLTM}(\text{State})$}
\State{$A' \leftarrow \text{getEquivalentAction}(\text{best action})$}
\If{$S' \notin \Mcal$}
\State{$\Mcal$.append($S'$)}
\EndIf
\If {best reward is significantly better for $A'$}
\State{$\Mcal[S'].\text{action} \leftarrow A'$}
\State{$\Mcal[S'].\text{reward} \leftarrow \text{best reward}$}
\EndIf
\EndFor
\end{algorithmic}
}
\end{algorithm}

(\textbf{Reading Dual-Memory: Getting Action})
Agents can retrieve an optimal action from the dual-memory for a state that was already visited. LTM is searched first, then STM is searched.
If an abstracted state is found to be in both, the action stored in LTM takes priority unless the reward for the action in STM is significantly higher. 

\begin{figure}[h]
    \centering
    \begin{subfigure}
        \centering
        \includegraphics[width=0.41\textwidth]{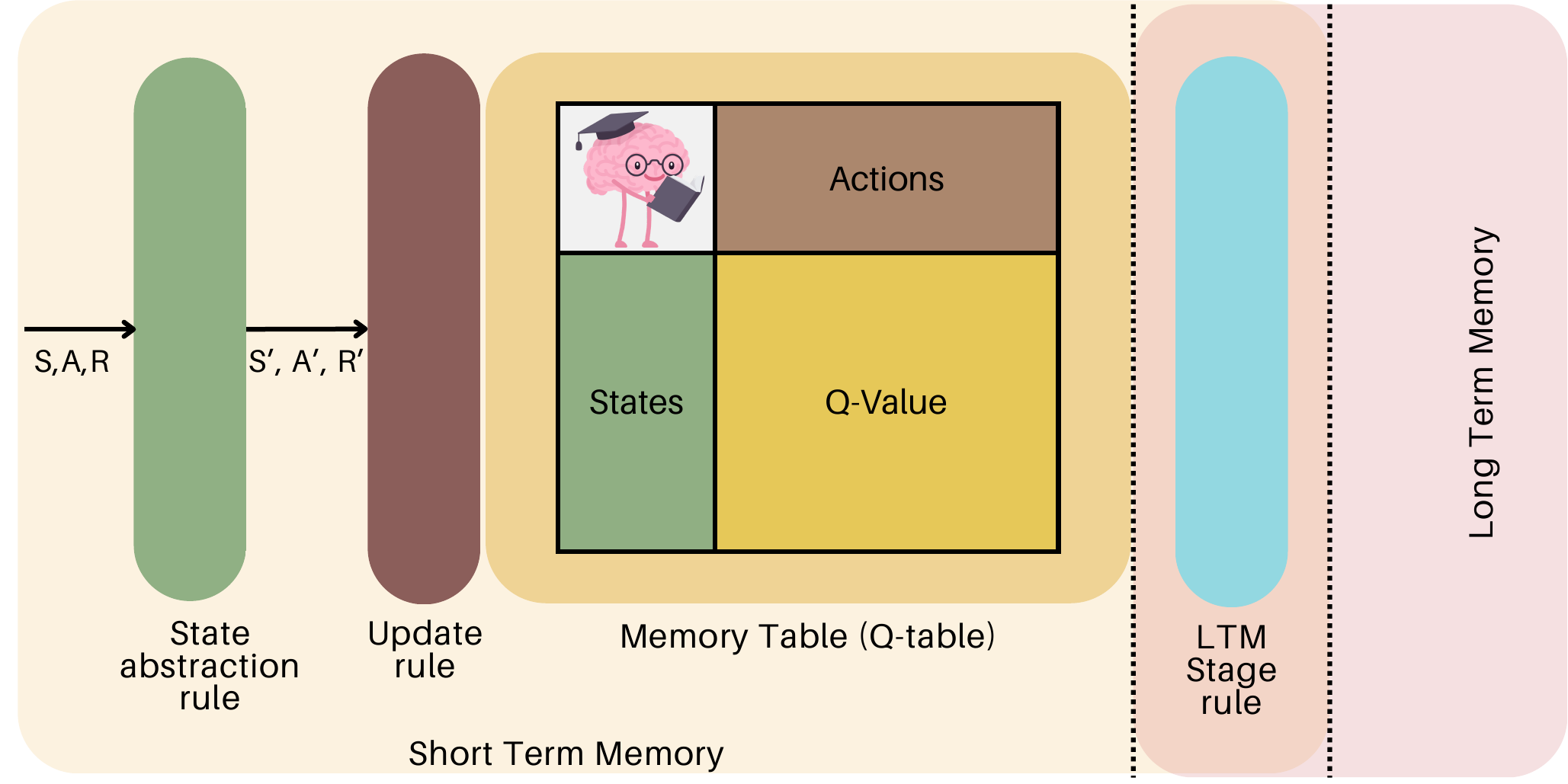}
        \label{fig:short_term_memory_arch}
    \end{subfigure}
    \hfill
    \begin{subfigure}
        \centering
         \includegraphics[width=0.4\textwidth]{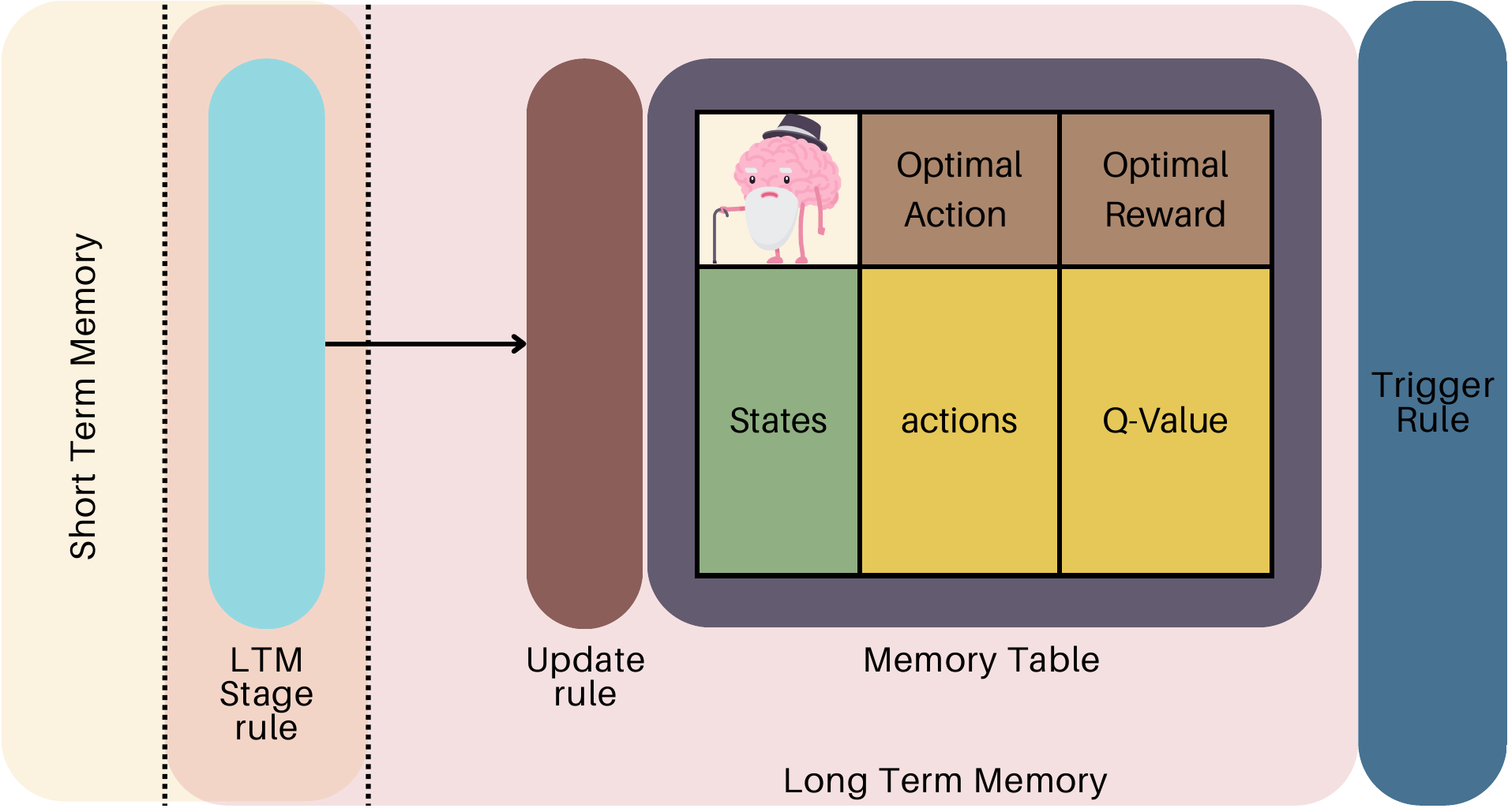}
        \label{fig:long_term_memory_arch}
    \end{subfigure}
    \caption{Architectures for STM (top) and LTM (bottom).}
    \label{fig:mem_arch}
    \vspace{-3pt}
\end{figure}
(\textbf{Memory-Communication Trigger Rule}) We employ a periodic trigger rule: after every $T$ seconds, each agent (intersection) shares its current memory (STM or LTM) with the CPU, which synchronises them to fill in any experience gaps. 
This CPU communicates with all agents in order to consolidate the experiences from the individual agents and promote more coherent learning of the environment.
The aggregation is done by averaging the Q-values to keep the scale consistent with the ones stored in each intersection. 

Our full memory system is implemented by~\alg{dmil}.

\begin{algorithm}
\small{
\caption{Dual-Memory Integrated Learning}
\label{alg:dmil}
\begin{algorithmic}[1]
\State{Initialise DualMemory}
\While{not done}
\State{$S_{\text{next}}, R, \text{done} \leftarrow$ environment.step()}
\State{$A_{\text{next}} \leftarrow $}DualMemory.$getAction()$ \Comment{$\epsilon-$greedy}
\State{$Q \leftarrow \text{getQValue}(S, A, R, S_{\text{next}}, A_{\text{next}})$}
\State{DualMemory.insert$(S, A, Q)$}
\If{include communication layer}
\\ \Comment{send data to CPU if triggered}
\State{DualMemory.$triggerRule()$} 
\EndIf
\EndWhile
\end{algorithmic}
}
\end{algorithm}


\subsection{Non-Memory vs. Memory-based Agents}\label{subsec:agent_types}
\begin{figure}[h]
    \centering
    \begin{subfigure}
        \centering
        \includegraphics[width=0.35\textwidth]{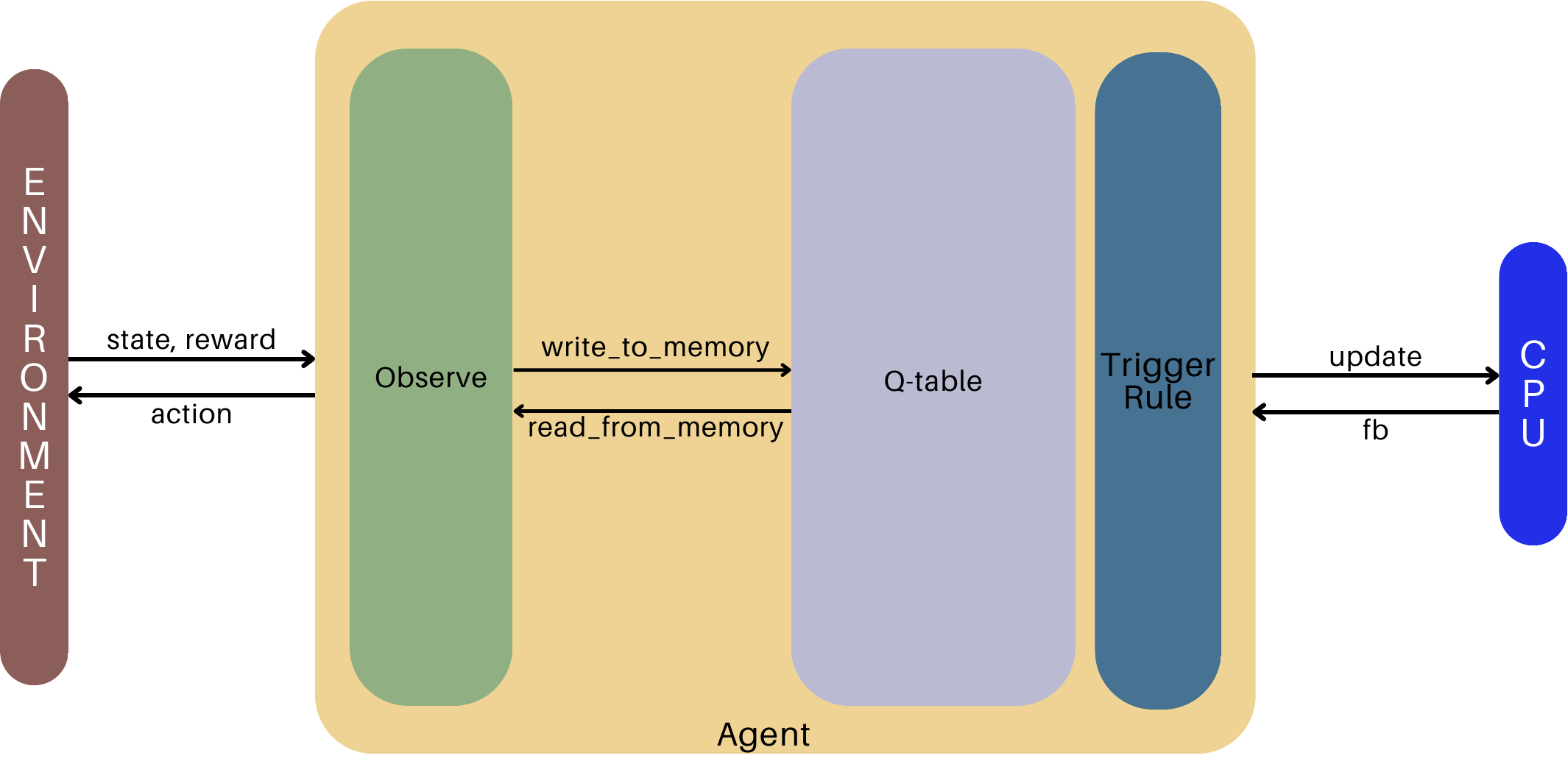}
        \label{fig:memory_less_agent_architecture}
    \end{subfigure}
    \hfill
    \begin{subfigure}
        \centering
         \includegraphics[width=0.46\textwidth]{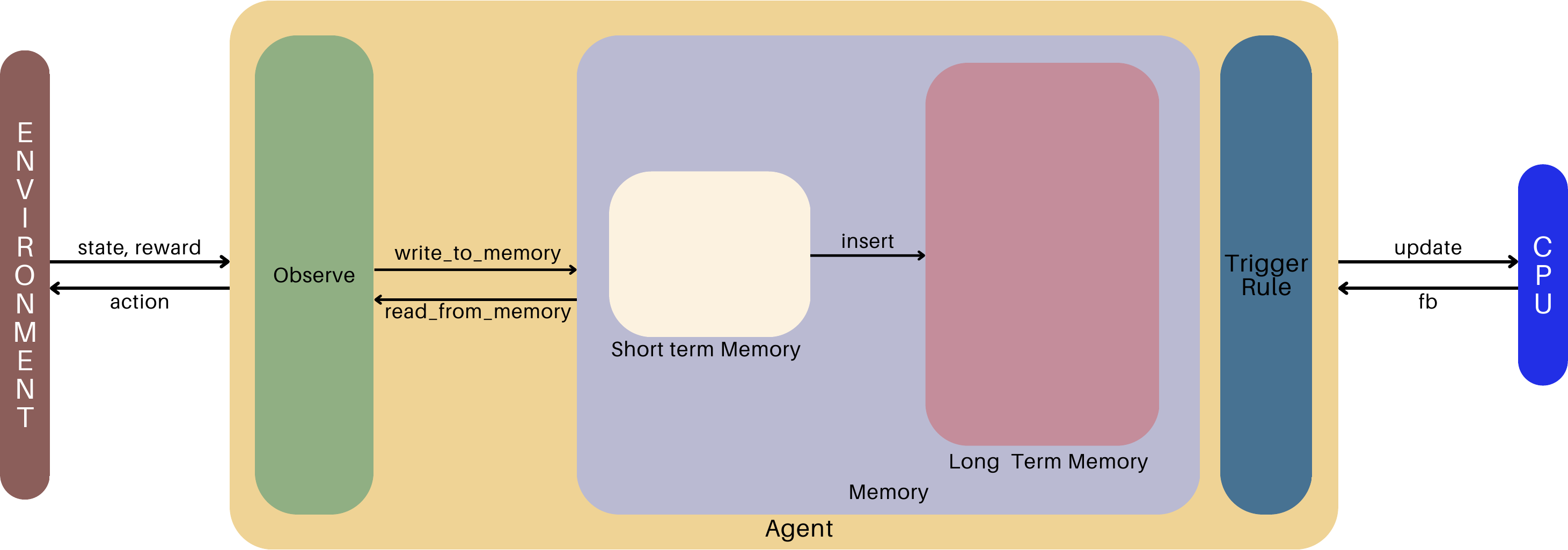}
        \label{fig:agent_architecture}
    \end{subfigure}
    \caption{Agent architectures for [top] SARSA (i.e., non-memory) and [bottom] memory-based.}
    \label{fig:agent_architectures}
    \vspace{-10pt}
\end{figure}

(\textbf{SARSA Agent}) The complete architecture of the SARSA agent can be seen in the top of Figure~\ref{fig:agent_architectures}.
It implements the standard SARSA model as described in~\cite{suttonbarto}, with Q-table update rule $Q(s_t,a_t){\,\leftarrow\,}Q(s_t,a_t) + \alpha[R_{t+1} + \gamma Q(s_{t+1},a_{t+1}) - Q(s_t,a_t)]$.
Because this agent does not include dual-memory, we may also refer to this as the ``non-memory'' agent.

(\textbf{Memory-based Agent}) The memory-based agent implements dual-memory integrated learning to keep track of the observed states. The agent implements a `forget' mechanism through the implementation of the STM and LTM structures. Similar to short-term and long-term memory in humans, the agent forgets the states that are encountered less and keeps only the most frequent states in the LTM. If the agent learns an action which yields better returns for an abstracted state which is already in memory (either in STM or LTM), the agent updates it.

\section{Theoretical Analysis}\label{sec:analysis}
We now analyze the growth bounds for the two types of agents defined in Section~\ref{subsec:agent_types}, in order to demonstrate the effectiveness of our dual-memory architecture.
In addition, our calculations allude to bounds on the size of action space, bounds on the rate of STM-to-LTM staging, etc., which help us tune our architecture's hyperparameters.

\begin{definition}\label{def:analysis_defs}
Let $T_{stage}{\,\in\,}\mathbb{N}$ be the period of the LTM Stage Rule.
%
%
Define $m_L[t]$ to be the number of states in the LTM at time $t$; this means the size of LTM is $3m_L[t]$ (counting three objects--state, action, and $Q$-value).
Define $m_s[t]{\,=\,}(m_s[t-1]{\,+\,}1)(1{\,-\,}I[t])$ to be the number of states in the STM, where 
$$
I[t]  = \begin{cases}
    1, & \text{if } t\mod{T_{stage}} = 0, \text{ and } t\neq 0 \\
    0, & \text{otherwise}
\end{cases}
$$
The size of STM at time $t$ is $|\mathcal{A}|m_s$ and we set $m_s[0]{\,=\,}1$.
The total size of the dual-memory architecture is then $msize_{Dual}[t]{\,\triangleq\,}3m_L[t] + |\mathcal{A}|m_s$.
For a SARSA agent, define $m_Q[t]$ to be the number of states in the replay buffer; the size of the agent's buffer at time $t$ is then $|\mathcal{A}|m_Q$.
Our comparison metrics are as follows.
Let $\kappa$ as the ratio of STM entries that are staged to the LTM.
Define the comparison ratio of the SARSA replay buffer to the dual-memory as $\zeta{\,\triangleq\,}msize_{Q}[t]/msize_{Dual}[t]$. Note $\zeta{\,=\,}1$ at $t{\,=\,}0$.
\end{definition}

Our main result (\lem{mem_size}) shows that the rate of growth of the dual-memory structure is always bounded above by a typical SARSA replay buffer.
This further suggests that dual-memory integrated learning is a more memory-efficient algorithm compared to more traditional RL methods with replay buffers.
A proof sketch of this result consists of tracing the worst and best-case trajectories of each memory architecture and computing their comparison ratios.

\begin{lemma}[Worst-Case Trajectory]\label{lem:worst_traj}
    Consider the worst-case scenario where all the states encountered until time $t\triangleq nT_{stage}$ are unique (i.e., the number of equivalence classes equals the number of states). 
    Assume $|\mathcal{A}|{\,\geq\,}3\kappa$.
    Then, in the worst case scenario where memory scales linearly, $msize_Q[t]\geq msize_{Dual}[t]$ i.e $\zeta\geq 1$.
\end{lemma}
%
Due to the nature of staging, $msize_{Dual}$ follows a staircase function with jumps at $t=nT_{stage}$. Hence to prove \cref{lem:worst_traj} it suffices to analyse the behavior at $t=nT_{stage}$ and $t=nT_{stage}-1$, i.e. before and after the LTM staging. From construction we have 
$m_L[t]{\,=\,}m_L[t-1] + I[t](m_s[t-1]\kappa)$, with $m_L[0]= 0$.
For $n>0$, observe that $m_s[nT_{\text{stage}}-1] = T_{\text{stage}}$. 
\begin{align*}
\text{At } t &= nT_{stage}: & \zeta_1 &= \frac{(nT_{stage}+1)|\mathcal{A}|}{|\mathcal{A}|+n3\kappa T_{stage}}\\
\text{At } t &= nT_{stage} - 1: & \zeta_2 &= \frac{(nT_{stage})|\mathcal{A}|}{T_{stage}|\mathcal{A}|+(n-1)3\kappa T_{stage}}
\end{align*}
For $\zeta_1\geq 1$ and $\zeta_2\geq 1$, we get $|\mathcal{A}|\geq 3\kappa$. This implies that the size of the action space sets an upper bound on the staging ratio of STM to LTM, and is relevant for small action space as $0<\kappa<1$.
For $\zeta_1\geq\zeta_2$, we get $|\mathcal{A}|\geq 3\kappa (T_{stage}+2)/(T_{stage}-2)>0 \implies T_{stage}>2$. Hence,  $msize_Q[t]\geq msize_{Dual}[t]$ for all $t$ in the worst-case scenario.

\begin{lemma}[Best-Case Trajectory]\label{lem:best_traj}
    Consider the best-case scenario where after some time $t{\,>\,}\tau$, each state encountered is placed inside an existing equivalence class. Let $M$ be the number of all unique encountered states. Then at any time $t{\,=\,}\tau + nT_{stage}$ with $n{\,>\,}0$,  $m_L[t] = M$. Moreover, under the same assumptions as~\cref{lem:worst_traj},
    $msize_Q[t]\geq msize_{Dual}[t]$ i.e $\zeta\geq 1$. 
\end{lemma}
%
The proof of~\lem{best_traj} follows directly from construction.
\begin{align*}
\text{At } t &= \tau + nT_{\text{stage}}: & \zeta_1 &= \frac{M \times |\mathcal{A}|}{|\mathcal{A}| + 3M}\\
\text{At } t &= \tau + nT_{\text{stage}}-1: & \zeta_2 &= \frac{M \times |\mathcal{A}|}{T_{stage}|\mathcal{A}| + 3M}
\end{align*}
For $\zeta_1{\,\geq\,}1$, $M{\,\geq\,} |\mathcal{A}|/(|\mathcal{A}|-3)$ is required. As $M{\,>\,}0 {\,\implies\,}|\mathcal{A}|{\,>\,}3$. \cref{lem:worst_traj} still holds.
For $\zeta_2\geq1$, $M{\,\geq\,} T_{stage} |\mathcal{A}|/(|\mathcal{A}|-3)$ is required. As $|\mathcal{A}|/(|\mathcal{A}|-3)$ is bounded below by 1 in $|\mathcal{A}|$, $M{\,>\,}T_{stage}>2$. It can also be quickly seen that $\zeta_1{\,\geq\,}\zeta_2$ is still satisfied. Hence, both $\zeta{\,\geq\,}1$ and \cref{lem:best_traj} holds.

\begin{lemma}[Comparison Ratio of Dual-Memory and SARSA]\label{lem:mem_size}
    At any time $t$, the total size of the dual-memory is less than or equal to the size of the SARSA replay buffer.
\end{lemma}
The proof of this follows directly by observing that $msize_{\text{dual}}$ is always less than or equal to $msize_{\text{Q}}$ in both the worst-case trajectory (showed by~\lem{worst_traj}) and the best-case trajectory (showed by~\lem{best_traj}).
Moreover, any time staging happens, the size of LTM increases at most by $\kappa m_s$ states, as opposed to the SARSA replay buffer which increases with each new state encountered.
Thus, at any time and for any trajectory between the best and worst-case scenarios, the dual-memory structure has less than or equal to the number of entries in the Q-table.

\begin{remark}[Optimal Hyperparameter Design]
    Dual-memory integrated learning introduced a collection of new hyperparameters (e.g., $\kappa$, $T_{\text{stage}}$, etc.) which can be tuned for more optimal performance. For \cref{lem:mem_size} to hold, we require $|A|{\,>\,}3$ and $M{\,\geq\,} T_{stage} |\mathcal{A}|/(|\mathcal{A}|-3)$ from the environment; and dual memory hyper parameters to satisfy $\kappa \leq |\mathcal{A}|/3$, and $T_{\text{stage}}{\,>\,}2$.
    Decreasing $\kappa$ increases the $\zeta$, but it also reduces the new long-term experiences learned by the agent.
    Increasing $T_{\text{stage}}$ decreases the rate of convergence of $\zeta_2$ to $\zeta_1$, however, it also increases the minimum number of unique states $M$ required for dual-memory to be more efficient in all scenarios.
\end{remark}


\section{Experiments}\label{sec:experiments}
We benchmark the performance of our dual-memory integrated learning framework on several network grids: single-intersection ($1\times 1$), $3\times 3$, and $5\times 5$. 
Incoming vehicles are modeled as a homogeneous Poisson process, parameterized with arrival rates $\lambda_D{\,>\,}0$ for each direction $D$; the specific values used are $\lambda_D{\,\in\,}\{5, 5, 10, 7\}$.
We emphasize that Poisson processes are chosen here for simplicity of experiment design, but our proposed approach can be applied to more complex scenarios, including real-world traffic data. 
Each vehicle takes the shortest path from one fringe node to another, with its destination node generated uniformly at random with the starting node excluded.
We choose $t_a{\,=\,}10$, $\theta{\,=\,}10$, $\alpha{\,=\,}1.1$, $\gamma{\,=\,}0.95$, $\epsilon_{\text{train}}{\,=\,}0.1$, $\epsilon_{\text{memory-based}}{\,=\,}0.4$, $\epsilon_{\text{non-memory}}{\,=\,}0.6$. 
The $\epsilon$ hyperparameters were obtained via grid search.

\subsection{Different Architecture Designs}\label{subsec:compare_architecture}
For comparison, we implement several different versions of dual-memory integrated learning based on the different equivalence class embeddings from~\subsec{equivalence}.
Additionally, we consider several modifications to the basic architecture described in the paper so far.

\textbf{(Communication Layer)}
We add a \textit{communication layer} to the CPU and agent network in the original setting of~\sec{problem_formulation}.
This is different from \textit{independent} MARL, where each agent optimizes its own local expected reward independently of the neighboring agents. 
With the communication layer, all agents' memories are synchronised, similar in vein to value decomposition network (VDN)~\cite{sunehag2017valuedecomposition}. However, in our case, the stored value function in memory is a function of abstracted states and actions. Furthermore, we assume that the value function of the whole system is the average of all agents' value functions (Remark~\ref{rmk:marl_avg}).
One caveat is that communication does not necessarily improve global optimality \cite{dewitt2020independent}, and this is a phenomenon we'll observe in our numerical simulations.

\textbf{(Reward Function Design)}
In addition to the basic reward function described in \cref{sec:problem_formulation}, we consider two other reward functions, entropy and diffusion, inspired by principles from physics.
We consider slight modifications to the original metrics in order to retain magnitude and sign information, which are important to distinguish traffic flow.\\
%
(\textit{Diffusion}) We treat an intersection as a medium for vehicles (particles) to pass through. This system aims to maximise the diffusion, calculated by the mean signed squared displacement of the particles, $-\text{mean}(\text{sign}(d)d^2)$, where $d$ is the difference between the next state and the current state.\\
%
(\textit{Entropy}) The entropy for our system is calculated by the difference $w_{k+1}{\,-\,}w_{k}$ between current and next state weighted entropy $w_{k}{\,\triangleq\,}- \sum_j(s_{k_j}\ln(s_{k_j}))$.
Similar to diffusion, the entropy of the whole network should be high, as low entropy implies congestion in a lane. 

The table of components combined to create the different dual memory architectures are summarized in~\tab{compare_architectures}.
The different combinations are created by choosing one entry from each column (for a total of $24$ implementations).
Independent SARSA agents with the default reward are used as the baseline for comparison against the other architectures.
\vspace{-.2cm}

\bgroup
\def\arraystretch{1.5}
\begin{table}[H]
    \begin{adjustbox}{width=\columnwidth,center}
      \begin{tabular}{| c | c | c | c |}
        \hline
        \textbf{MARL Type} & \textbf{Agent} & \textbf{Equiv. Class (\subsec{equivalence}}) & \textbf{Reward} \\ \hline \hline
        Independent & Non-Memory (SARSA) & Simple & Default (\sec{problem_formulation}) \\ \hline
        Communicating & Memory (memory-based) & Complex & Diffusion\\ \hline
        -- & -- & -- & Entropy\\ \hline
      \end{tabular}
    \end{adjustbox}
    \caption{}
    \vspace{-15pt}
    \label{tab:compare_architectures}
\end{table}
\egroup

\subsection{Training and Testing Pipeline}\label{subsec:pipeline}
Each architecture from~\subsec{compare_architecture} is fed through the following training and testing pipeline.\\
(\textbf{Training}) 
A dummy SARSA agent is trained offline for $T_{\text{sim}}$ timesteps of incoming vehicles, gathered from data.
The Q-table from this SARSA agent is fed into all the agents we are simulating.
Here, our simple dual-memory agent refers to the agent which implements dual-memory with a simple equivalence class that only utilizes shift invariance.
Next, our complex dual-memory agent refers to the one that utilizes all three symmetries described in Section~\ref{subsec:equivalence}.
Finally, we have a baseline SARSA agent used to benchmark both our both simple and complex agents.
We remark that the dummy SARSA agent is simply used to accumulate prior knowledge; this enables continuous comparison since all agents start from the same prior knowledge.\\ 
(\textbf{Testing}) As mentioned, the agents are given the trained Q-table from the dummy SARSA agent in order to start everyone from a common baseline and maintain fair testing. 
The performance metrics we consider are described in the following Section~\ref{subsec:metrics}.

\begin{remark}
    We did not add any distinct operations behind creating the Q-table for both the dummy and baseline SARSA agents. 
    Both agents perform the usual SARSA update described in Section~\ref{subsec:agent_types}, and fetch the best state-action pair when not exploring.
\end{remark}

\subsection{Evaluation Metrics}\label{subsec:metrics}
To compare the proposed methods, we used three different evaluation metrics. 
First is the number of vehicles arriving at their destinations ($V_c$), which is calculated by counting every vehicle that has reached its randomly assigned destination. For the single intersection, $V_c$ is the same as the number of vehicles crossing the intersection. Our second metric is the average waiting time ($W$) of all vehicles in the network. We normalized the number with $V_c|_{t=300}$ to properly compare the algorithms. Finally, to compare the different equivalence embeddings, we record the memory size ($m_L$) evolution throughout the simulation. For the SARSA agent, we evaluate the Q-table size instead.

\subsection{Results and Discussion}
%


\begin{figure*}[h]
\centering 
\begin{subfigure}
    \centering
    \includegraphics[width = 0.4\textwidth]{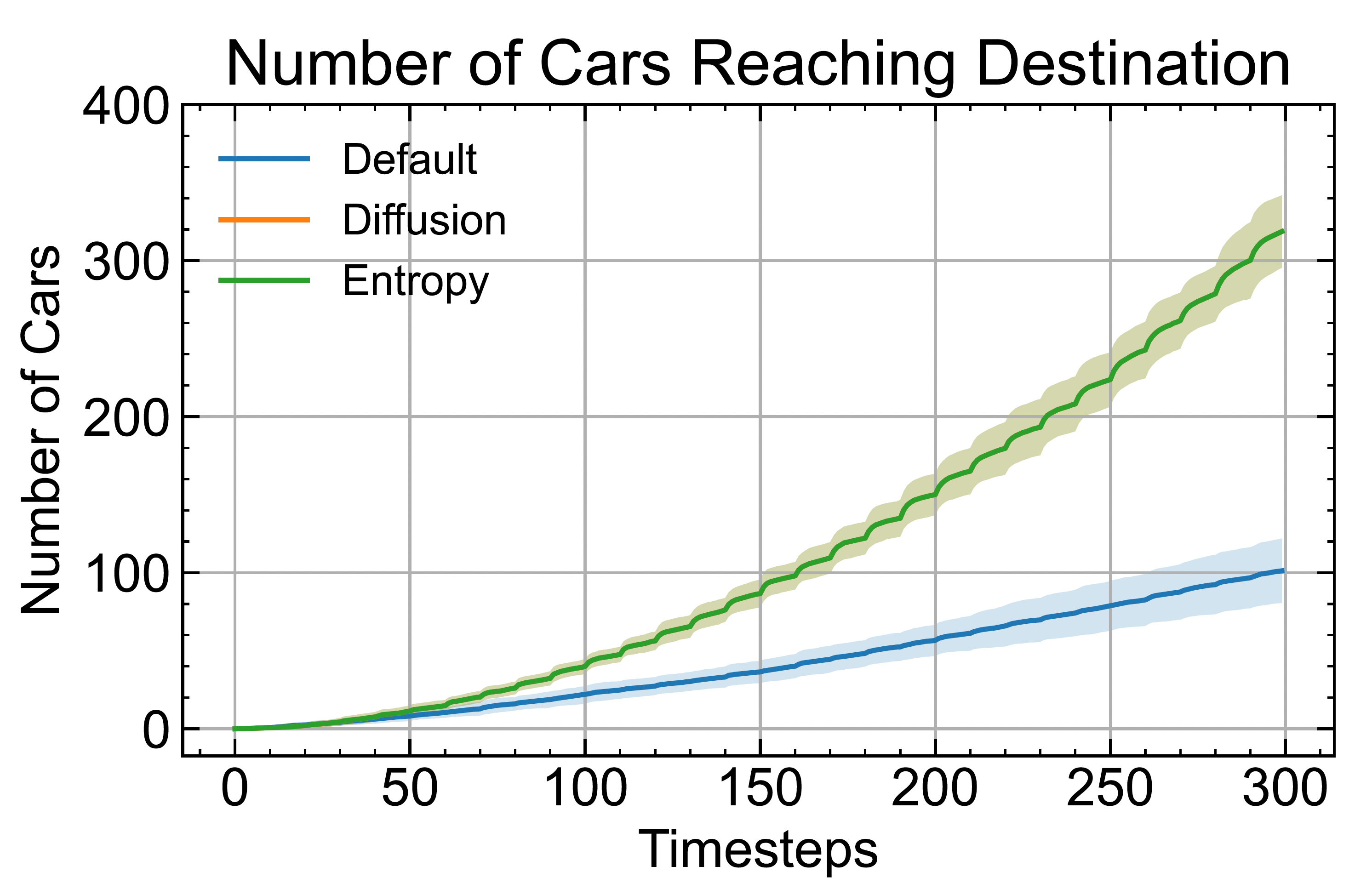}
    \label{fig:vc_memory}
\end{subfigure}
\hspace{.3cm}
\begin{subfigure}
    \centering
    \includegraphics[width = 0.4\textwidth]{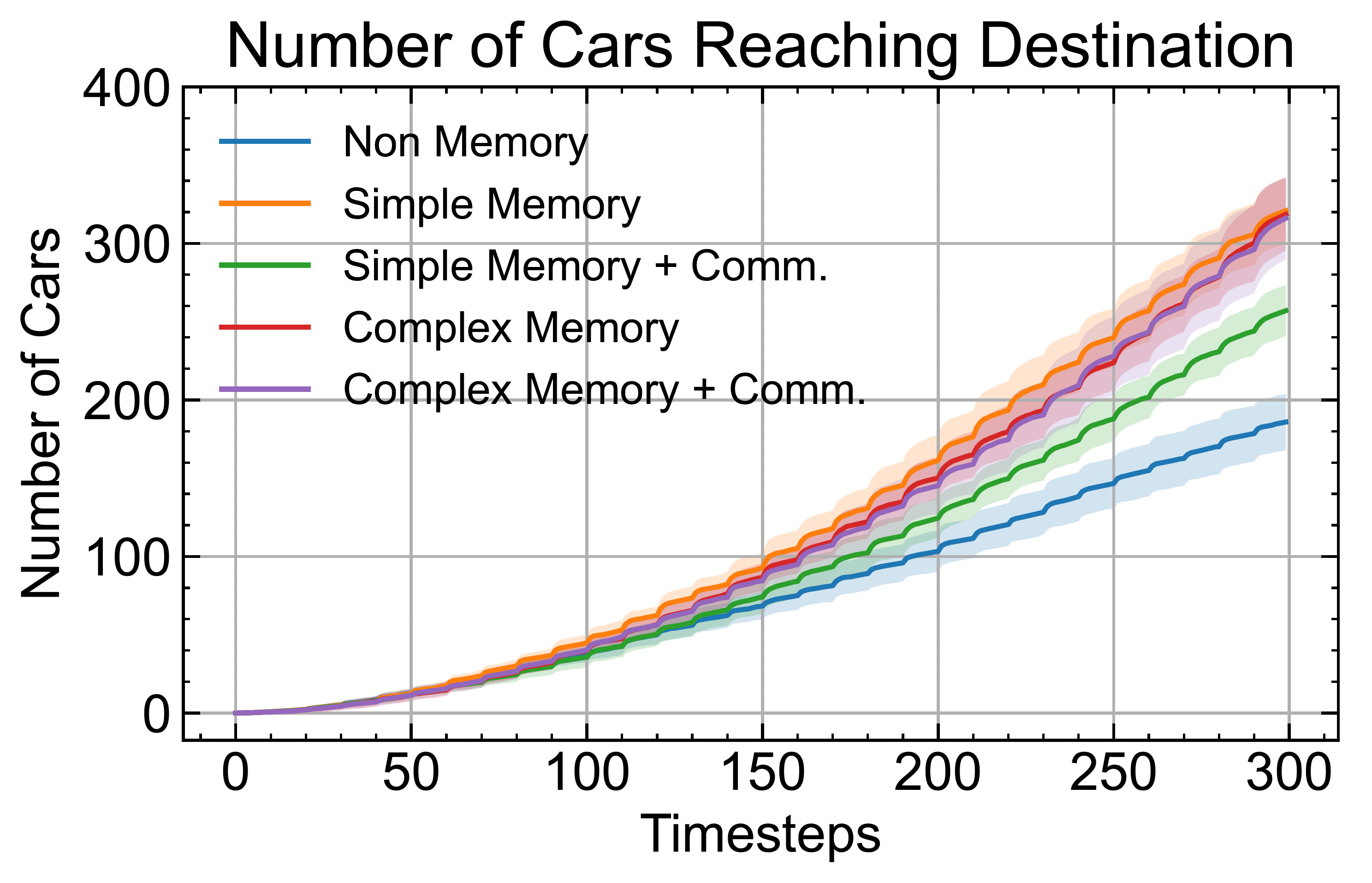}
    \label{fig:arrivi}
\end{subfigure}
\caption{Number of vehicles reaching destination ($V_c$) for the $5\times 5$ grid network.
[Left] A comparison of the performance among different reward functions. [Right] A comparison among different memory architecture types, with entropy reward fixed.
}
\label{fig:vc_comparison}
\vspace{-10pt}
\end{figure*}

All simulations were run on an Intel Core i3-10100 machine with 32GB RAM and were simulated on 30 different random seeds. 
The average waiting time ($W$) comparison between a SARSA agent and a memory-based agent in a single intersection can be seen in Table~\ref{tab:w_comparison_normalized}. 
The memory-based agent with simple and complex equivalence methods is labeled with ``Simple'' and ``Complex'', respectively. ``Complex + Communication'' indicates the case where the communication layer is used to synchronize the memories with complex equivalent methods.
In the single-intersection case, the dual-memory agent has a longer average waiting time than the SARSA agent. 
However, for both the $3\times 3$ and $5\times 5$ networks, memory-based agents perform better than the SARSA agents. 
The improvement arises mainly because there are many more unseen states for each agent in the multiple intersection setting. 
Relying solely on the SARSA algorithm is not enough to return the optimal actions since the learning requires state visitation. 
Using equivalence classes allows the agent to infer the best action for states without ever seeing them before.


The performance of intersection control can be enhanced further if different reward functions are used. Using entropy and diffusion decreased $W$ up to 30\% (Table \ref{tab:w_comparison_normalized} and increased the vehicle throughput by more than 100\% (Figure~\ref{fig:vc_comparison}). An interesting result here is that the largest improvement from different reward functions occurs in the memory with the complex equivalence embedding. This occurs because the method coalesces the scale-invariant states into one, which lets the agents efficiently choose the action that could maximize the spread of the vehicles. This behavior is amplified further by the entropy or diffusion reward, showcasing the importance of choosing the right reward signal for a particular environment.

\begin{table}[h]
    \vspace{-5pt}
    \centering
    \resizebox{\columnwidth}{!}{%
    \begin{tabular}{c|c|c|c|c|c}
        Grid Type & Reward & No Memory & Simple & Complex & Complex + Communication\\
        \hline
        \hline   
        \multirow{3}{2em}{1x1} & Default & 0.21 $\pm$ 0.1 & 0.24 $\pm$ 0.1 & 0.29 $\pm$ 0.1 & -\\
        & Entropy & 0.14 $\pm$ 0.1 & 0.16 $\pm$ 0.1 & 0.17 $\pm$ 0.1 & -\\
        & Diffusion & 0.14 $\pm$ 0.1 & 0.16 $\pm$ 0.1 & 0.17 $\pm$ 0.1 & -\\
        \hline
        \hline        
        \multirow{3}{2em}{3x3} & Default & 0.19 $\pm$ 0.1 & 0.16 $\pm$ 0.03 & 0.24 $\pm$ 0.05 & 0.25 $\pm$ 0.05\\
        & Entropy & 0.1 $\pm$ 0.02 & 0.08 $\pm$ 0.01 & 0.08 $\pm$ 0.01 & 0.08 $\pm$ 0.01\\
        & Diffusion & 0.1 $\pm$ 0.02 & 0.08 $\pm$ 0.01 & 0.08 $\pm$ 0.01 & 0.08 $\pm$ 0.01\\
        \hline
        \hline
        \multirow{3}{2em}{5x5} & Default & 0.17 $\pm$ 0.04 & 0.16 $\pm$  0.03 & 0.26 $\pm$ 0.04 & 0.24 $\pm$ 0.04\\
        & Entropy & 0.08 $\pm$ 0.01 & 0.05 $\pm$ 0.004 & 0.06 $\pm$ 0.004 & 0.06 $\pm$ 0.003\\
        & Diffusion & 0.08 $\pm$ 0.01 & 0.05 $\pm$ 0.004 & 0.06 $\pm$ 0.004 & 0.06 $\pm$ 0.003
    \end{tabular}
    }
    \caption{}
    \label{tab:w_comparison_normalized}
     \vspace{-15pt}
\end{table}

Another noticeable point from Table~\ref{tab:w_comparison_normalized} and {Figure~\ref{fig:vc_comparison}} is that entropy and diffusion reward yield nearly identical $V_c$ performance. These signify that the traffic congestion reinforcement learning is not sensitive to reward signal scaling, unlike other environments such as Atari games where agents learn differently depending on how we scale the reward signals~\cite{haarnoja_soft_2018}.
We also observe that introducing communication often increased waiting times in the $3\times 3$ and $5\times 5$ grid settings with the default reward function ({Figure~\ref{fig:vc_comparison}).
Several references in the literature (e.g.,~\cite{dewitt2020independent}) suggest that for some tasks, dependent MARL agents perform worse than independent MARL agents because the added dependency among agents forces the agents to exhibit similar Q-values even if the same action may not optimal for some agents.
This leads to problems such as \textit{relative overgeneralization}, where the solution converges to a suboptimal joint action.
Table~\ref{tab:w_comparison_normalized} shows that this phenomenon doesn't occur with the other two reward signals.

The comparison of the memory sizes can be seen in Figure~\ref{fig:memory_size_comparison} for $5{\,\times\,}5$ network (the $3{\,\times\,}3$ case followed similar trends, and so is omitted). 
Unlike the SARSA Q-tables, which increase each timestep, dual-memory learning agents exhibit flatter slopes, indicating a slower growth rate. By abstracting the states with the equivalence classes, the number of memory entries are greatly reduced, while having similar or even better congestion metrics. Furthermore, the addition of communication and even the simple equivalence embedding method reduces the memory size further with a tiny congestion performance decrease.

\section{Conclusion and Future Work}\label{sec:conclusion}
Dual-memory integrated learning, which uses a short-term, long-term two-phase memory table with equivalence classes allows for a more sample-efficient implementation of MARL, especially for traffic management.
We showed through analysis and experiments that memory growth rate is reduced with little expense on the performance.
Different architecture combinations and additional engineering (e.g., communication layer, reward design) may yield additional performance gains.

There are several avenues of future work.
The performance of dual-memory integrated learning on more complex networks can be investigated.
Further analysis on the communication layer, such as synchronization delay effects and the inclusion of predictions based on neighboring agent information~\cite{pmlr-v211-han23a}, can be done.
Advanced triggering and staging rules will also be designed.
Other methods of equivalence class embeddings will also be explored. 

\bibliographystyle{IEEEtran}
\bibliography{%
main%
}

\begin{figure}
\centering 
    \includegraphics[width = 0.45\textwidth]{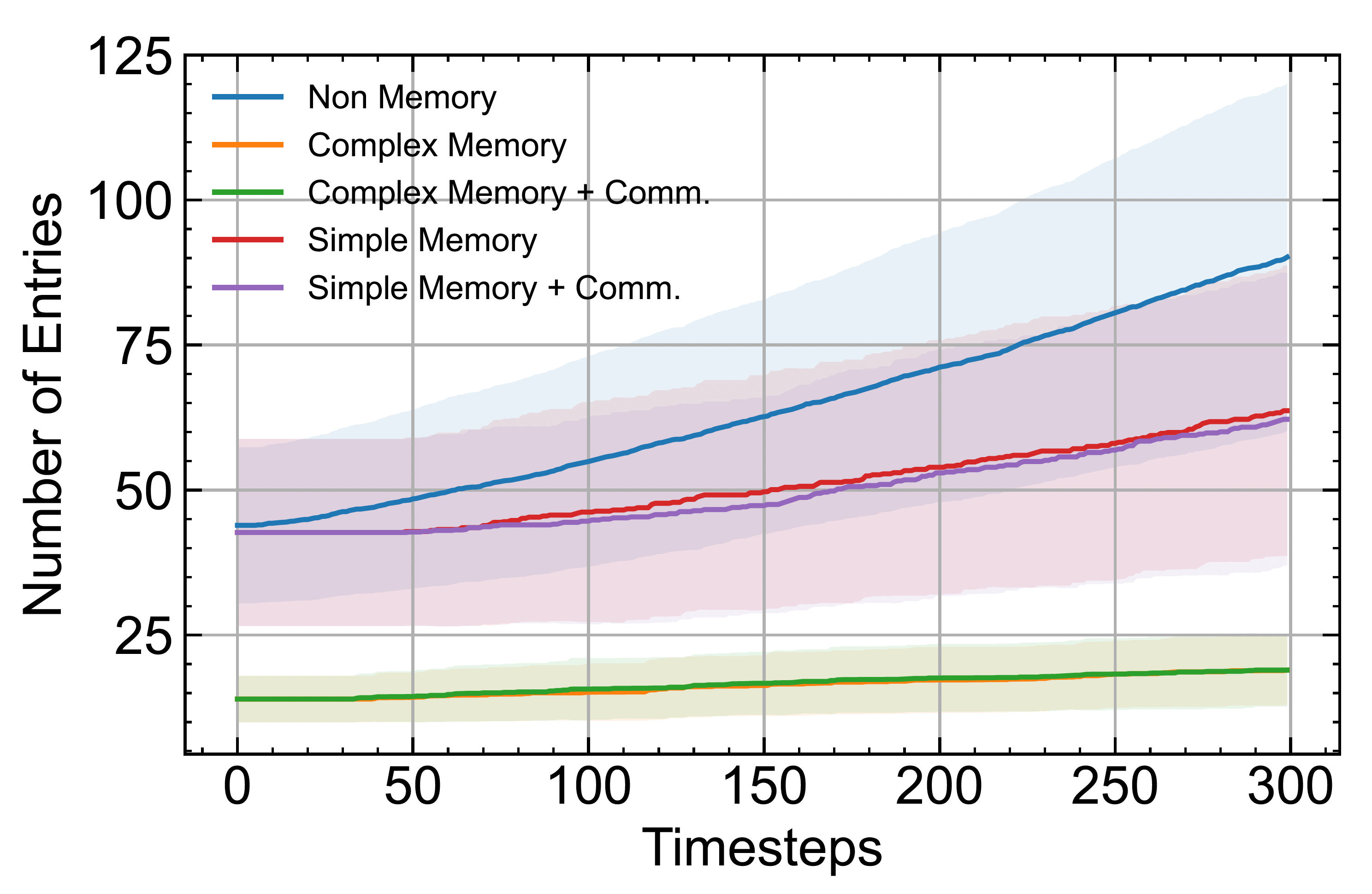}
    \vspace{-5pt}
    \caption{Memory table growth vs. time for the $5\times 5$ network, averaged over all intersections. For SARSA (non-memory), the number of Q-table entries are shown instead. 
    }
    \label{fig:memory_size_comparison}
    \vspace{4.6in}
\end{figure}

\end{document}